# WatchOut: A Road Safety Extension for Pedestrians on a Public Windshield Display

**Matthias Geiger, Changkun Ou, Cedric Quintes**
University of Munich (LMU)
Munich, Germany
{firstname.lastname@campus.lmu.de}

**ABSTRACT**
We conducted a field study to investigate whether public windshield displays are applicable as an additional interactive digital road safety warning sign. We focused on investigating the acceptance and usability of our novel public windshield display and its potential use for future applications. The study has shown that users are open-minded to the idea of an extraverted windshield display regardless the use case, whether it is used for safety purposes or different content. Contrary to our hypothesis most people assumed they would mistrust the system if it were as well established as traffic lights and primarily rely on their own perception.

**ACM Classification Keywords**
H.5.2 Information Interfaces and Presentation: User Interfaces

## INTRODUCTION
In this work, we investigated, if a vehicle's extraverted windshield display can improve the awareness of pedestrians to the traffic situation when crossing the street. We conducted a field study using a car with a windshield display showing a safety visualization for pedestrians to signal them whether a vehicle is approaching and in consequence for the pedestrian whether it is safe to cross the road. The study has shown that the approach is a novelty, which pedestrians showed strong interest in. Users were creative when imagining safety use cases and overall embraced the idea of the establishment of the public windshield display for various reasons explained in the course of this paper. The field study has shown that the vast majority of pedestrians was not aware of the display due to its novelty. Most people just have never seen a windshield display before and thus they did not realize it at first glance. When they got aware of the display they showed a vigor interest in the visualization and the intended purpose. Overall, they have seen the display and intended visualization as supportive and useful for traffic situations under the condition that it is well-tested and established in everyday traffic - to improve traffic safety people have to know the system to be able to use it. When the message of the display (see Figure 2) has been well understood, some pedestrians were irritated by the icon coding which is mostly related to the novelty of the approach. Unanimously, pedestrians came to the conclusion that the display has the potential to improve the safety in traffic situations but would have to be further tested and established in traffic.

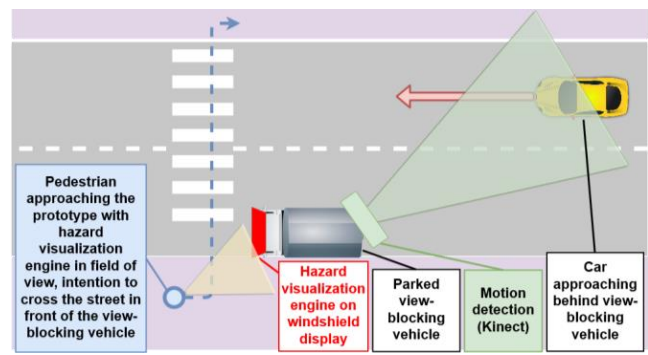

**Figure 1. Hazard visualization for pedestrian approaching a view-blocking vehicle with a car approaching behind it**

## RELATED WORK
In the last decade, smartphones gradually established a deep involvement in people's everyday life. Due to the omnipresence of displays and their variety (tablet, smartphone, PC, laptop, etc.) people nowadays are accustomed to the presence of displays all around, even in cars. Car displays are usually used to display information directed to users inside the car. Co-drivers can use displays for entertainment and drivers get visual feedback via the CID. First implementations of windshield displays (WSDs) [1] extend the CID by using the windshield as a display to e.g. visualize traffic-relevant [2], navigation-related information [3] or contents for entertainment [4]. The awareness on public interactive displays has been investigated in a study in 2012 [5], which has shown that significantly more passers-by tend to notice interactivity late and have to walk back to interact. If somebody is already interacting, others begin interacting as well (honeypot effect). Displaying information explicitly to the outside of the car is practically unchartered territory. Possible use-cases for the application of outward displays [6] were described to provide useful information like empty parking lots, commercials entertainment and mentioned a use case for safety improvements by warning by-passing cars about e.g. scary turns, bumps and pot holes.

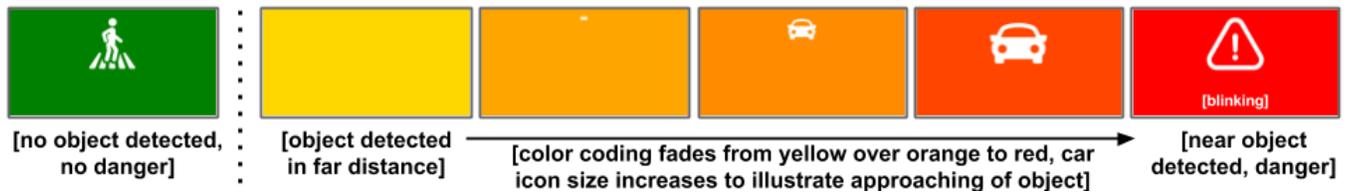

**Figure 2. Hazard Visualization**

## CASE STUDY

Based on a framework for evaluating public displays [7] we conducted a descriptive field study. In this context, we were interested in the installation's aptitude for increasing road safety, pedestrians' behavior, user experience, acceptance and the effectiveness of the display with a focus on validity of the results for evaluating the windshield display. Furthermore, we used a design space classification [1] to define the prototype composition. The prototype is designed as a road safety enhancement within a parked car, which is observed by multiple road users. The purpose of the visualization was raising awareness among pedestrians about potential danger situations, when cars are approaching to a possible street crossing location behind the installation. The visualization on the windshield is registered in 2D within the observer's periphery and shows a symbolic presentation that varies in color, size and motion. Both, the display brightness and the ambience illumination, are primary factors influencing the visualization quality.

### Hazard Visualization

The hazard visualization (see Figure 2) relies on colors and symbols, which were evaluated in a brief pre-study and are well known among public from traffic signs by intuition according to two focus groups consisting of three persons each, whom we showed different icons and variations in color coding to evaluate their salience and understandability in order to find the most appropriate design.

While no approaching object is detected behind the parked view-blocking vehicle, the windshield display shows a green background with an icon of a pedestrian walking on a crosswalk in „secure mode". As soon as an approaching object is detected, the visualization switches to "hazard mode", starting with a light orange background color and showing a small car icon. We defined segments a range of 25 meters, with 5 parts – one for 5 meters distance. The closer the detected object approaches to the street crossing position of the pedestrian, the larger the car icon is scaled and the more the background color fades darker orange tones. In the nearest detection situation with the greatest hazard for pedestrians, the visualization shows a blinking exclamation mark sign with a signal-red background color. Altogether, the part "hazard mode" of the visualization consists of five increments. The hazard visualization was implemented as a web application based on HTML5, CSS3, JavaScript and jQuery and in any state-of-the-art browser. The different increments of the visualization are triggered from a motion detection engine, or manually switched by an operator.

### Motion Detection Tests

A motion detection engine was built based on a Microsoft XBOX Kinect sensor (first generation) and a custom driver, which is implemented in Python and uses the provided depth image of the sensor to detect approaching objects and their current range in real-time. During the preliminary tests, the motion detection worked properly for objects in up to approx. eight meters distance inside a closed room with artificial illumination. In a real-life environment, we observed, that the detection quality and range heavily decreases in daylight conditions. On the one hand, too bright illumination, e.g. at full sunshine, and on the other hand, a dark setting, e.g. from evening to sundown, showed negative effects on our detection results. Consequently, we decided to manually control the hazard visualization in a Wizard-of-Oz-study.

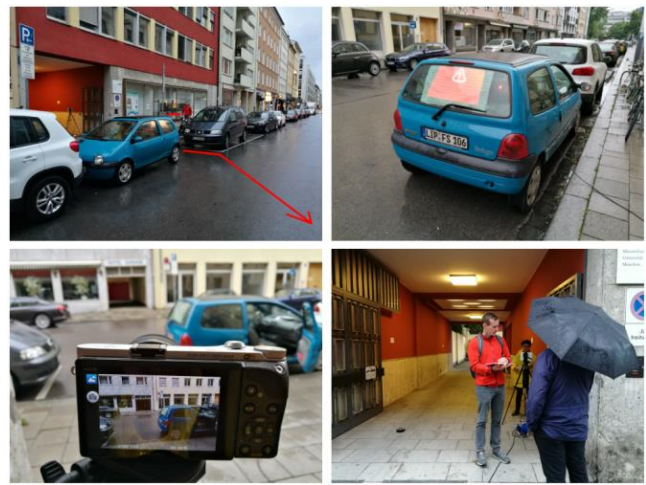

**Figure 3. Experimental setup for the field study**

### Experimental Setup

For the descriptive field study, a Renault Twingo was parked at a location on the side of a street, where people tend to cross due to prior observation. The rear windshield of the car was covered with half-transparent white paper that exhibits a low opacity. Using this surface, it is possible to display content on it using back-projection from a projector inside the car (see Fig. 3, top-right). The projector is connected to a laptop, which runs the hazard visualization within Google Chrome. Due to the insufficient motion detection, a team member was hiding inside the car and controlling the hazard visualization manually, as soon as an object approached in direction of the street crossing spot. Since the public awareness of the display installation depends on its perceptibility and high illumination it is crucial to compare the influence of the environment lightning. To overcome illumination bias, the study in two time frames - daytime (6pm - 8pm); nighttime (9pm - 11pm) - in the summer with occasional drizzle and a stable, saunter-friendly weather condition.

As seen on Figure 1, an optimal scenario for the study would be a location where pedestrians tend to cross the street and with high levels of traffic. The location for the study was chosen by the following criteria: An on average constant stream of people approach the installation, at day time and night time. Furthermore, the installation shall not provoke any safety issues since the focus of the study is to examine the users' awareness of the display.

**Subjects and Measurements**

With observation, video recording and interviews of pedestrians surrounding the windshield display, we collected quantitative and qualitative data about user interaction, increase of road safety, acceptance and use cases to assess the effectiveness of the content displayed on the screen as well as the social impact regarding the analysis of reactions. For a neutral measurement of observation and to verify the data from the questionnaire and the study conductors' observations, a video camera was used during the interviews. Passing pedestrians were surveyed with a questionnaire to gather qualitative data about understanding, acceptance and usage of a public windshield display as an additional road safety warning. In order to verify the pedestrians' statements each pedestrian was observed and all answers were logged on a printout of the questionnaire by the interviewer.

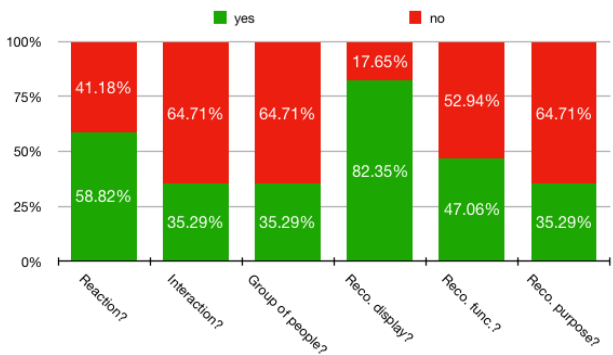

**Figure 4. Quantitative Analysis Results**

## DATA ANALYSIS

**Participants**

The user study was conducted in two periods within a day from 6pm to 8pm during daytime and from 9pm to 11pm during nighttime. We recruited 17 subjects in total, each subject either an individual person (7 male, 4 female) or a group of acquaintance (N=6, Mean=2.33). Among all participants, 10 subjects during daytime and 7 subjects during night time. The daytime subjects have 4 groups of acquaintance (Mean=2.25) and 6 individuals (4 male, 2 female); the nighttime subjects have 2 groups of acquaintance (Mean=2.5) and 5 individuals (3 male, 2 female). All participants answering the questionnaire were pedestrians with the obvious intention to cross the street.

**Quantitative Analysis**

The statistical results (see Figure 4) indicate that 58.82% of the pedestrians showed a reaction to the windshield display; 61.28% of the pedestrians with reactions performed interactions with the display; 50.0% of pedestrians who showed interactions approached in a group. Pedestrians whose attention has been drawn by the display in general first stopped and after a short pause they walked straight up to the display to inspect it in detail. The second part of qualitative results provides insights into the pedestrians' subjective opinions regarding the usability of the windshield display.

35.3% of the pedestrians understood the intended purpose of the visualization. The pedestrians received the interpretation correct at their first glance. 82.35% of pedestrians recognize the display, and 47.06% of the pedestrians recognized the display functionalities. More precisely, the participants assessed the installation as a feature for road safety and most of them were close to the original purpose of a warning sign at a possible street crossing location for pedestrians.

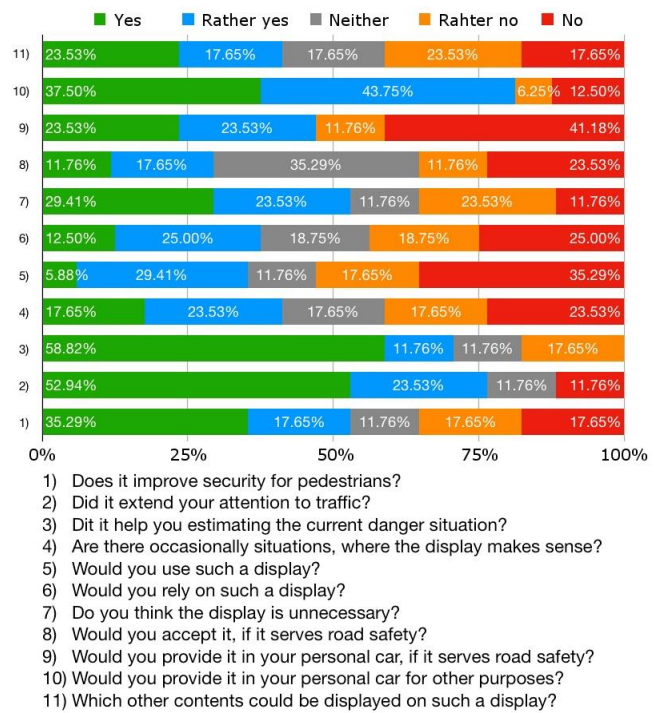

1) Does it improve security for pedestrians?
2) Did it extend your attention to traffic?
3) Dit it help you estimating the current danger situation?
4) Are there occasionally situations, where the display makes sense?
5) Would you use such a display?
6) Would you rely on such a display?
7) Do you think the display is unnecessary?
8) Would you accept it, if it serves road safety?
9) Would you provide it in your personal car, if it serves road safety?
10) Would you provide it in your personal car for other purposes?
11) Which other contents could be displayed on such a display?

**Figure 5. Qualitative analysis results and related questions**

Figure 5 shows the Likert scale of the pedestrians' subjective perspective on the usability of the prototype windshield display. The participants have a neutral opinion about the installation's safety and doubt its estimation accuracy. Most of the participants (81.25%) are accepting this installation and would be willing to accept it in their personal cars if it works well (64.71%). Among all subjects, we found three interesting results through applying significance tests:

1) **Day-time vs. night-time**: We assume that as the null hypothesis H0 for the nighttime group safety does not significantly improve. Using the Mann-Whitney-U test the result (U=177.0, p=0.107>0.050, accept H0) indicates that nighttime subjects and daytime subjects do not show any significant difference regarding the level of perceived rad safety. The result proves that the installation' environment (bright or dark lightning situation) doesn't influence the pedestrians' attitude towards perceived road safety.

2) **Pedestrians interests**: In the hypothesis H0 we assume, that people do not show a significant difference in the purpose of using the installation for safety reasons vs. other reasons. We performed a One-way ANOVA test. The result (F=2.50, p=0.124>0.050, accept H0) indicates that people do not show any significant difference in the purpose of using the installation for safety reasons vs. other reasons. Considering the acceptance (81.25%) of the pedestrians and the significance test, we conclude that the pedestrians showed a vigor interest and well accepted the windshield display for their cars.

3) **Conditional reliability**: Finally, considering that people who accept the installation for safety reasons would significantly rely on the installation as H0 hypothesis, we performed a One-way ANOVA test and the result (F=12.25, p=0.001<0.050, reject H0) indicates, that people who accepted the installation for safety do not significantly rely on the installation. More precisely, people tend to not completely rely on the installation, even if they accept it. We will discuss the reason for this phenomenon it in the subsequent section of safety, based on qualitative analysis.

**Qualitative Analysis**

**Safety:** The technology has to be established and thereby proven to be safe and useful, but pedestrians would primarily rely on their own senses. Most people who were questioned see the display as a security risk for pedestrians as long as the new technology is barely known in public and not well tested. Furthermore, some people were concerned about the distraction of pedestrians through the salience of the novel technology. The blinking illuminated display and the color coding fostered peoples' awareness on the traffic situation - especially during night time - and made people aware of the traffic situation. Main concerns were technical issues triggering malfunctions and user's opinions were two-fold regarding the necessity of the display.

**Acceptance:** Some participants stated that the existing traffic lights are sufficient, some agreed that the installation increases security next to view-blocking cars. All participants stated that they would accept it, in case that it improves road safety and works reliably. Most participants would provide the display in their own car if it is free or offered as a standard feature, but would relinquish displaying arbitrary content.

**Privacy:** Reasons for refusing the installation were privacy issues and the indignation of providing personal information in public. In contrary, some people would like the idea of earning money through advertisement on the display. Following other use cases and contents were mentioned as possible contents: traffic news, nearby public transportation connections, nearby objects of interest, news & social media feeds, advertisement, personal messages to other pedestrians, movies.

**DISCUSSION**

Despite the successful quantitative and qualitative analysis, the study design contains a few drawbacks, which might influence our results. Because the prototype is installed in a single car, we are not able to project our conclusion on the situation of universally installed (pervasive) windshield displays in a widespread range of cars. In case of ubiquitous appearance, the road safety could either improve, as the technology gets well known among pedestrians, but the distraction of pedestrians could also increase due to the larger amount content available. Our study was conducted in two time slots (daytime and nighttime). 82.36% of the pedestrians were aware of the display among daytime and nighttime. However, the windshield display becomes more salient at night time, when the ambient light is lower. For future works, an alternative visualization should replace the back-projection from inside the car. Furthermore, a proper motion detection system, whose accuracy is independent of ambient light, should be used for detecting approaching cars.

**CONCLUSION**

Most people confronted with the display in our field experiment showed a reaction. Due to the novelty of the approach most users were confused at first glance regarding the use of the display but showed a vigor interest in the course of the interview. They predominantly thought of the display as a potential security improvement to traffic situations while they had difficulties to imagine it as a security risk as long as the technology is tested and well established. Since the display is a novel prototype that is not established in traffic situations yet people showed a low level of trust in the system and would rather prefer to solely rely on their own senses when crossing the street. If the system would be well tested and the technology established most could imagine relying on the display as well. People unanimously found the display to be useful and showed a high level of acceptance for the display in their environment. The vast majority would permissively provide a display in their personal car under the condition that it is free of charge and promotes the overall safety in traffic situations. In our field study, we focused on the pedestrian's attention and awareness of the display and the correlated security indications. In a future work, one may gradually shift the focus from the measurement of awareness to the measurement of security and potential security issues. Furthermore, the display could be A/B-tested with a multitude of different traffic situations to identify scenarios where and if the display is helpful for pedestrians and if the use is in fact related to specific traffic scenarios or not.

**ACKNOWLEDGMENTS**

Thanks to our supervisors Renate Haeuslschmid, Dr. Bastian Pfleging and Prof. Dr. Florian Alt for their support and guidance during the project.

**REFERENCES**

1. Haeuslschmid, R., Pfleging, B., & Alt, F. (2016, May). A design space to support the development of windshield applications for the car. In Proceedings of the 2016 CHI Conference on Human Factors in Computing Systems (pp. 5076-5091). ACM.
2. Haeuslschmid, R., Schnurr, L., Wagner, J., & Butz, A. (2015, September). Contact-analog warnings on windshield displays promote monitoring the road scene. In Proceedings of the 7th International Conference on Automotive User Interfaces and Interactive Vehicular Applications (pp. 64-71). ACM.
3. Kim, S., & Dey, A. K. (2009, April). Simulated augmented reality windshield display as a cognitive mapping aid for elder driver navigation. In Proceedings of the SIGCHI Conference on Human Factors in Computing Systems (pp. 133-142). ACM.
4. Kim, M. J., Yoon, S. H., & Ji, Y. G. (2016, January). Exploring the User Experience for Autonomous Vehicle and the Role of Windshield Display: Based on Framework Approach. In Proceedings of HCI Korea (pp. 321-326). Hanbit Media, Inc.
5. Müller, J., Walter, R., Bailly, G., Nischt, M., and Alt, F. Looking glass: A field study on noticing interactivity of a shop window.
6. Selker, T., Burleson, W., & Arroyo, E. (2002, April). E-windshield: a study of using. In CHI'02 Extended Abstracts on Human Factors in Computing Systems (pp. 508-509). ACM.
7. Alt, F., Schneegaß, S., Schmidt, A., Müller, J., & Memarovic, N. (2012, June). How to evaluate public displays. In Proceedings of the 2012 International Symposium on Pervasive Displays (p. 17). ACM.